\documentclass{ws-mplb}

\begin{document}

\markboth{Boretsky, Cohn, and Freericks}{Spin and pseudospin towers of the Hubbard model on a bipartite lattice}

%
\catchline{}{}{}{}{}
%

\title{SPIN AND PSEUDOSPIN TOWERS OF THE HUBBARD MODEL ON A BIPARTITE LATTICE}

\author{J. Z. Boretsky}

\address{Department of Physics, McGill University,\\
3600 rue University\\
Montreal QC, Canada H3A 2T8\\
jonathan.boretsky@mail.mcgill.ca}

\author{J. R. Cohn}

\address{Department of Physics, Georgetown University,\\
37$^{\rm th}$ and O Sts. NW, Washington, DC 20057 USA\\
jrc260@georgetown.edu}

\author{J. K. Freericks}

\address{Department of Physics, Georgetown University,\\
37$^{\rm th}$ and O Sts. NW, Washington, DC 20057 USA\\
james.freericks@georgetown.edu}

\maketitle

\begin{history}
\received{(Day Month Year)}
\revised{(Day Month Year)}
\end{history}

\begin{abstract}
In 1989, Lieb proved two theorems about the Hubbard model. One showed that the ground state of the attractive model
was a spin singlet state ($S=0$), was unique, and was positive definite. The other showed that the ground state of the repulsive model on a bipartite lattice at half-filling has a total spin given by $|(N_A-N_B)/2|$, corresponding to the difference of the number of lattice sites on the two sublattices divided by two. In the mid to late 1990's, Shen extended these proofs
to show that the pseudospin of the attractive model was minimal until the electron number equaled $2N_A$ where it became fixed at $J=|(N_A-N_B)/2|$ until the filling became $2N_B$, where it became minimal again. In addition, Shen showed that a spin tower exists for the spin eigenstates for the half-filled case on a bipartite lattice. The spin tower says the minimal energy state with spin $S$ is higher in energy than the minimal energy state with spin $S-1$ until we reach the ground-state spin given above. One long standing conjecture about this model remains, namely does the attractive model have such a spin tower for all fillings, which would then imply that the repulsive model has minimal pseudopsin in its ground state. While we do not prove this last conjecture, we provide a quick review of this previous work, provide a constructive proof of the pseudospin of the attractive model ground state, and describe the challenges with proving the remaining open conjecture.
\end{abstract}


\section{Introduction}

The single-band Hubbard model\cite{hubbard} involves electrons hopping on a lattice and their mutual on-site repulsion when up and down spin electrons occupy the same lattice site. It remains one of the most studied models in condensed matter physics, primarily because it has been solved exactly only in one-dimensions\cite{lieb_wu} and in infinite dimensions\cite{infinite-d}. One interesting problem which has yielded some powerful exact results involves the quantum numbers of the ground states in sectors with different electron number and different $z$-component of spin. In particular, the idea of a spin tower, where the minimal states with different spin are ordered according to the total spin eigenvalues, and a similar concept with respect to a pseudospin tower, remain interesting questions for this model. This work is partly motivated by the work by Lieb and Mattis in one-dimension\cite{lieb_mattis}, where they proved that a spin tower exists for one-dimensional models on a lattice. We argue below that it is likely that this result can be extended to all dimensions for the attractive case, but are not yet able to fully prove the result. Instead, we review much of the work of others on this problem\cite{lieb,shen,shen_review,lieb_freericks} and we show an alternative constructive proof for the existence of a spin tower for the repulsive Hubbard model on a bipartite lattice at half filling.

\section{Formalism}

We work with  single-band Hubbard model, which is defined by the following Hamiltonian on a general graph $\Lambda$ with $|\Lambda|$ vertices $x\in\Lambda$ (which we refer to as sites on the graph)
\begin{equation}
\mathcal{H}=-\sum_{x,y\in\Lambda, \sigma} t_{xy} c^\dagger_{x\sigma}c^{\phantom{\dagger}}_{y\sigma}
+U\sum_{x\in\Lambda}
c^\dagger_{x\uparrow}c^{\phantom{\dagger}}_{x\uparrow}
c^\dagger_{x\downarrow}c^{\phantom{\dagger}}_{x\downarrow}.
\label{eq: ham}
\end{equation}
The creation (annihilation) operators for a fermion of spin $\sigma$ at graph site $x$ are defined to be $c^\dagger_{x\sigma}$ ($c^{\phantom{\dagger}}_{x\sigma}$) and they satisfy the canonical anticommutation relation $\{c^\dagger_{x\sigma},c^{\phantom{\dagger}}_{y\sigma\prime}\}_+=\delta_{xy}\delta_{\sigma\sigma\prime}$
(with all creation operators anticommuting amongst themselves, and similarly for the annihilation operators). The matrix $-t_{xy}$ is the hopping matrix, which is required to be real and symmetric $t_{xy}=t_{yx}$ and to have a vanishing diagonal ($t_{xx}=0$) and $U$ is the 
on-site Hubbard interaction, which we will take to be the same for every lattice site. The attractive case corresponds to $U<0$, while the repulsive case corresponds to $U>0$.

\begin{figure}[th]
\centerline{\psfig{file=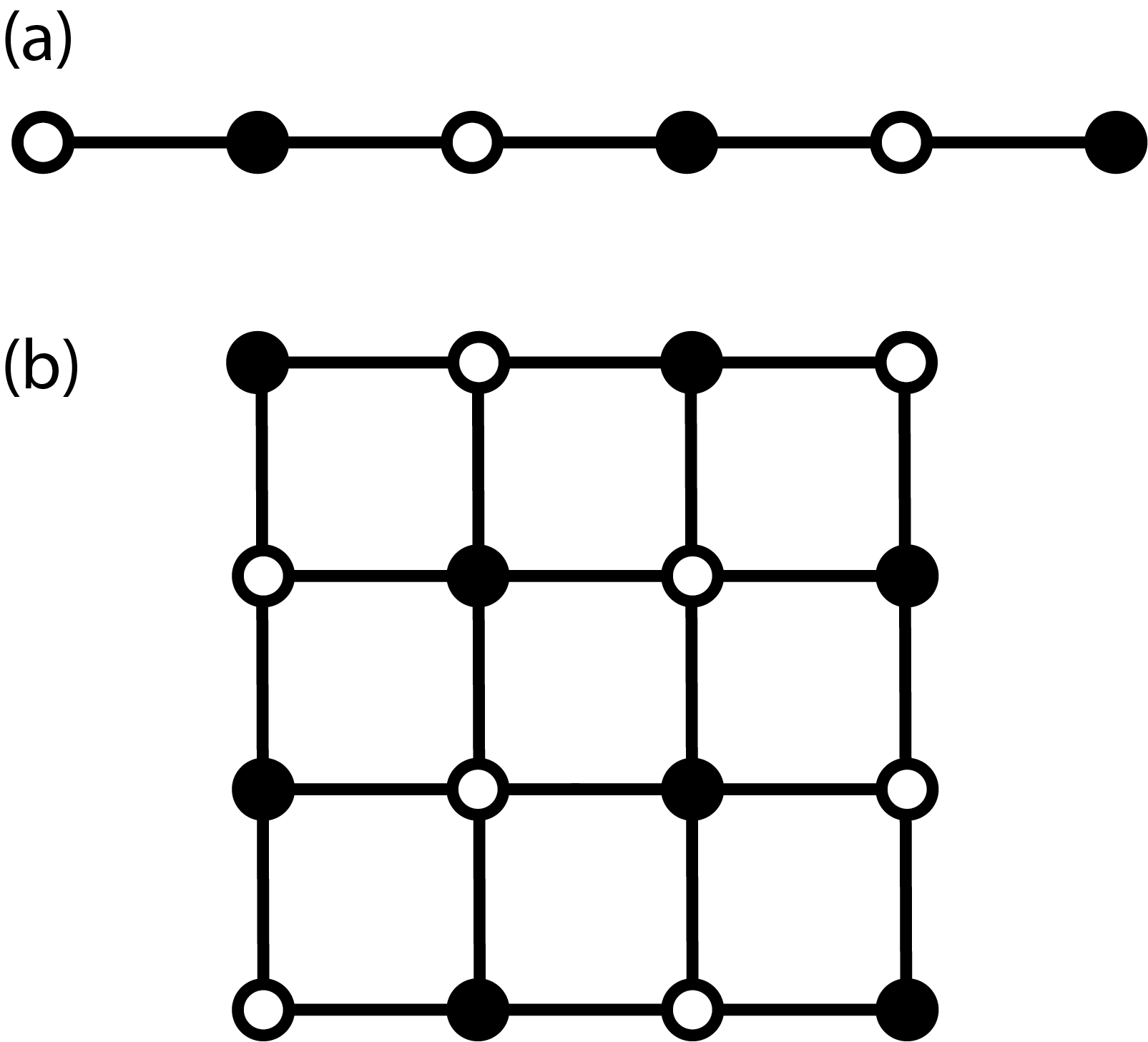,width=2.2in}~~~~~
\psfig{file=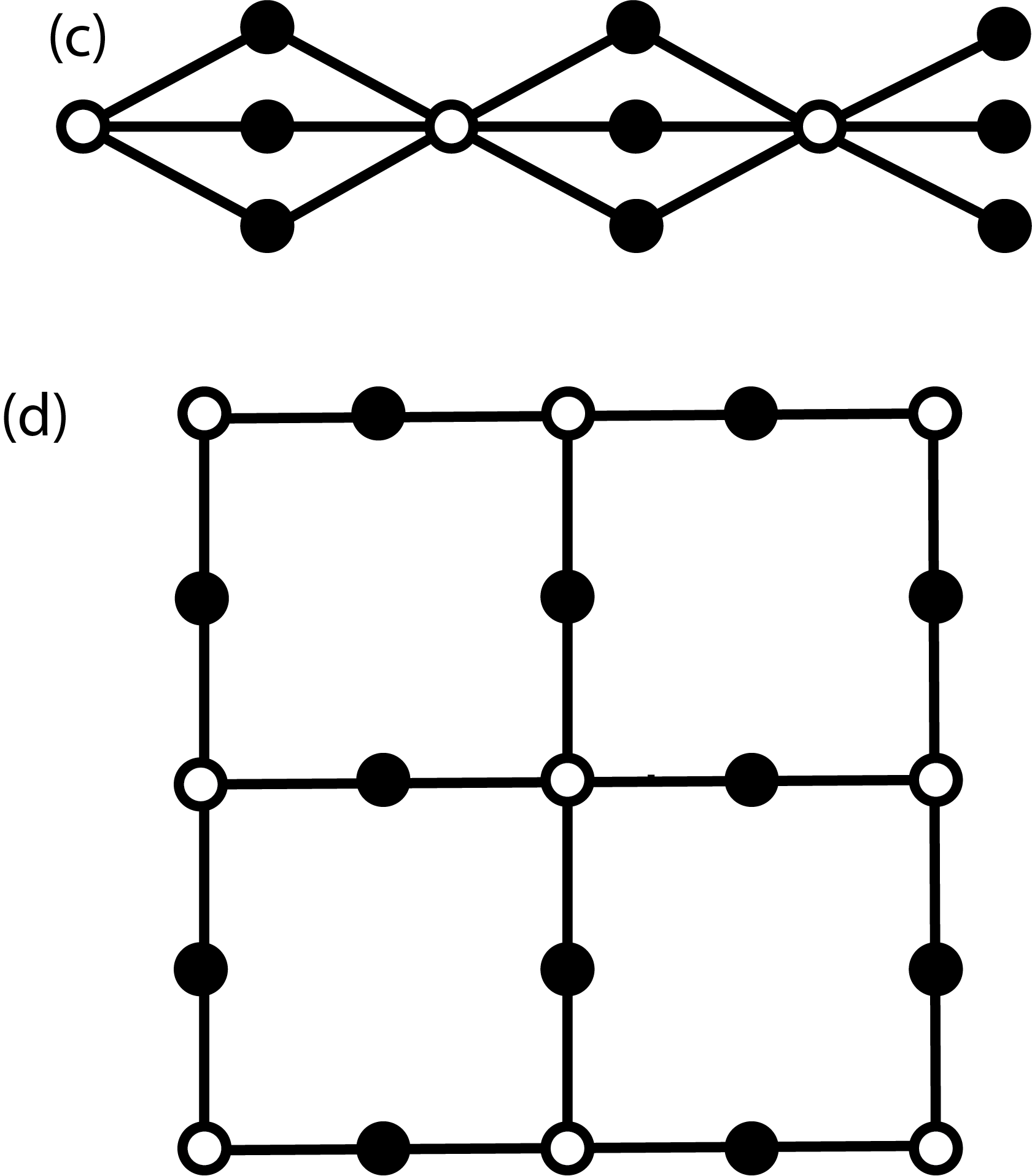,width=2.2in}}
\caption{Different bipartite lattice types. (a) One-dimensional linear chain ($N_A=N_B$) and (b) nearest-neighbor square lattice ($N_A=N_B$). (c) One-dimensional Lieb-lattice ($N_A\ne N_B$) and (d) two-dimensional Lieb-lattice ($N_A\ne N_B$). The open circles are the $A$ sublattice and the solid circles are the $B$ sublattice. The lines denote hopping matrix connections between lattice sites (which always connect solid circles to open circles). \label{fig: lattices}}
\end{figure}

In this work, we will focus solely on {\it bipartite lattices}. The graph is called a bipartite lattice if it separates into two pieces, one with $N_A$ lattice sites and one with $N_B=|\Lambda|-N_A$ lattices sites (we choose $N_A\le N_B$ here), and the hopping matrix $-t_{xy}$ is zero whenever $x~{\rm and}~y\in A~{\rm sublattice}$ and whenever 
$x~{\rm and}~y\in B~{\rm sublattice}$. In other words, all of the hopping is between the two sublattices ($x\in A$ sublattice and $y\in B$ sublattice, or {\it vice versa}). Note that we do not require periodicity, nor do we require $N_A=N_B$. It has become common to call a bipartite lattice where $N_A\ne N_B$ as a Lieb lattice. Examples of bipartite lattices with the same numbers of elements in each sublattice (all with nearest-neighbor hopping only) are the one-dimensional lattice, the square lattice, the simple cubic lattice, the body-centered-cubic lattice, and so on. See Fig.~\ref{fig: lattices} for some examples of conventional bipartite lattices and Lieb lattices.

One technical point we need to handle first is that the fermionic operators all anticommute with each other. We will assume without loss of generality that all fermionic operators with the same spin anticommute with each other, but those with different spins commute. This can be accomplished by making the unitary transformation $c^{\phantom\dagger}_{x\uparrow}\rightarrow (-1)^{N_{\downarrow}}c^{\phantom\dagger}_{x\uparrow}$ and 
$c^\dagger_{x\uparrow}\rightarrow (-1)^{N_{\downarrow}}c^\dagger_{x\uparrow}$. Here, we have $N_\sigma=\sum_{x\in\Lambda} n_{x\sigma}=\sum_{x\in\Lambda} c^\dagger_{x\sigma}c^{\phantom\dagger}_{x\sigma}$. The phase factor will change the anticommutator to a commutator whenever we anticommute two fermionic operators of different spin. As we will see below, because we work in a sector with fixed up and down spin particle numbers, this transformation has no effect on the energy eigenvalues, the quantum numbers, or the dynamics of the Hubbard model. So we simply assume that the same spin operators anticommute, but the opposite spin operators commute.

Our next step is to describe the operators that commute with the Hubbard Hamiltonian and to describe the operators whose quantum numbers we will be examining in the ground state and in the different energy towers that we consider in this work. The first set of operators we consider are the total number operators for spin up and spin down, which were defined above. We find that both $N_\uparrow$ and $N_\downarrow$ commute with $\mathcal{H}$ and hence the number of up spin particles and the number of down spin particles are separately conserved by the Hamiltonian. It turns out that we can construct two different $SU(2)$ algebras that either commute with the Hubbard Hamiltonian or they form raising and lowering operator relationships with $\mathcal{H}$. The first of these are the total spin $SU(2)$ operators, which are defined via
\begin{equation}
S^+=\sum_{x\in\Lambda}c^\dagger_{x\uparrow}c^{\phantom\dagger}_{x\downarrow}\,,\,\,\,
S^-=\sum_{x\in\Lambda}c^\dagger_{x\downarrow}c^{\phantom\dagger}_{x\uparrow}\,,\,\,\,{\rm and}\,\,\,
S^z=\frac{1}{2}(N_\uparrow-N_\downarrow).
\label{eq: spin_def}
\end{equation} 
These operators can be verified to satisfy the conventional $SU(2)$ algebra:
\begin{equation}
[S^z,S^\pm]_-=\pm S^\pm\,,\,\,\,[S^+,S^-]_-=2S^z
\label{eq: su2_spin}
\end{equation}
and they also all commute with $\mathcal{H}$: $[S^z,\mathcal{H}]_-=[S^\pm,\mathcal{H}]_-=0$. This means the eigenstates of the Hubbard Hamiltonian form total spin multiplets, with each member of the multiplet having the same energy.

Before we define the pseudospin operators, we must define the symbol $\Pi_x$ which is equal to 1 if $x\in A$ sublattice and equals 0 if $x\in B$ sublattice. Then the pseudospin operators become
\begin{equation}
J^+=\sum_{x\in\Lambda}(-1)^{\Pi_x}c^\dagger_{x\uparrow}c^\dagger_{x\downarrow}\,,\,\,\, J^-=\sum_{x\in\Lambda}(-1)^{\Pi_x}c^{\phantom\dagger}_{x\uparrow}c^{\phantom\dagger}_{x\downarrow}\,,\,\,\,{\rm and}\,\,J^z=\frac{1}{2}(N_\uparrow+N_\downarrow-|\Lambda|).
\label{eq: pseudo}
\end{equation} 

They also satisfy an $SU(2)$ algebra: $[J^z,J^\pm]_-=\pm J^\pm$ and $[J^+,J^-]_-=2J^z$. Furthermore, all spin operators and pseudospin operators commute. Finally, while the $z$-component of pseudospin commutes with the Hubbard Hamiltonian $[J^z,\mathcal{H}]_-=0$, the pseudospin raising and lowering operators are raising and lowering operators for $\mathcal{H}$: $[\mathcal{H},J^\pm]_-=\pm UJ^\pm$. So, as the eigenvalue of $J^z$ changes, the number of electrons in the state changes, and the energy of the eigenstate also shifts by multiples of $U$. This will become clear as we discuss some examples below when we discuss pseudospin towers.

The original proof by Lieb\cite{lieb} involves a variational argument with the wavefunction expressed in a form that is called the {\it spin-reflection symmetric form}. We start by fixing the values for the total electron number $N$ and the number of up spins $M$ and the number of down spins $N-M$. When considering the problem of an even number of electrons, we can search for the ground state in the sector where the eigenvalue of $S^z$ is equal to zero ($N=2M$), since every spin multiplet has a representative with $S^z=0$. We choose a complete set of $M$-particle basis functions for spinless fermions on the lattice denoted $\{\phi_\alpha\}$. These basis functions are chosen to be degree-$M$ polynomials of the creation operators with real coefficients. A simple counting argument shows there are exactly $|\Lambda|!/[M!(|\Lambda|-M)!]=R$ such wavefunctions for each spin. We write an arbitrary wavefunction with $2M$ electrons and $S^z=0$ as follows:
\begin{equation}
\psi=\sum_{\alpha,\beta=1}^{R} W_{\alpha\beta} \phi_{\alpha\uparrow}\otimes\phi_{\beta\downarrow},
\label{eq: w}
\end{equation}
where the $R^2$ complex numbers that determine the wavefunction are organized as an $R\times R$ matrix denoted by $W_{\alpha\beta}$. Note that the same basis functions are used for both spins and the order of the basis functions is the same for both spins as well.

The wavefunction is normalized, which implies that ${\rm Tr}W^\dagger W={\rm Tr}WW^\dagger =\sum_{\alpha\beta}|W_{\alpha\beta}|^2=1$. We define a many-body kinetic-energy matrix (for a single spin species) via
\begin{equation}
K_{\alpha\beta}=\langle \phi_\alpha|\left (-\sum_{x,y\in\Lambda }t_{xy} c^\dagger_{x}c^{\phantom\dagger}_{y}\right )|\phi_\beta\rangle
\label{eq: kinetic}
\end{equation}
and a many-body number matrix (also for a single spin species) via
\begin{equation}
(L_x)_{\alpha\beta}=\langle \phi_\alpha|c^\dagger_{x}c^{\phantom\dagger}_{x}|\phi_\beta\rangle .
\label{eq: n}
\end{equation}
Both of these matrices are Hermitian, of course, but because the basis functions are real and the same for both spins, these matrices are actually {\it real symmetric} matrices.

The Schr\"odinger equation $\mathcal{H}\psi=E\psi$ then becomes
\begin{equation}
KW+WK^T+U\sum_{x\in \Lambda}L_xWL_x^T=EW\,\,{\rm or}\,\,KW+WK+U\sum_{x\in \Lambda}L_xWL_x=EW
\label{eq: schroedinger}
\end{equation}
where matrix multiplication is implied and we used the fact that $K$ and $L_x$ are symmetric matrices (this is why we need the hopping matrix to be real). If we take the Hermitian conjugate of the latter equation, we find
\begin{equation}
W^\dagger K+KW^\dagger +U\sum_{x\in\Lambda} L_x W^\dagger L_x=EW^\dagger
\label{eq: schroedinger2}
\end{equation}
where we used the Hermiticity of the $K$ and $L_x$ matrices. This implies that if $W$ is a wavefunction with energy $E$, then $W^\dagger$ is also a wavefunction with energy $E$. By forming $W+W^\dagger$ and $i(W-W^\dagger)$, we can always find a wavefunction with a Hermitian matrix and the same energy $E$, so without any loss of generality, we can assume the $W$ is a Hermitian matrix (but this does not imply it is a real symmetric matrix, although it could be). Note that this proof shows that $W$ is Hermitian in the original real basis $\{\phi_\alpha\}$. But, because Hermiticity is preserved under all unitary transformations, this implies that $W$ is Hermitian in any basis that is a unitary transformation of the original basis.

\section{Summary of Lieb's proofs}

Lieb's proof that the ground state contains a spin-singlet state employs the following variational argument\cite{lieb}.
Using the fact that $W$ is Hermitian, allows us to write the nomalization condition as ${\rm Tr}W^2=1$ and we find the energy satisfies
\begin{equation}
E=2{\rm Tr}KW^2+U\sum_{x\in\Lambda}{\rm Tr}WL_xWL_x
\label{eq: energy}
\end{equation}
where we assume the wavefunction is normalized. We next define the matrix absolute value to satisfy
\begin{equation}
|W|=\sqrt{W^2}
\label{eq: sqrt}
\end{equation}
where we always choose the nonnegative square roots. This is constructed by first diagonalizing the matrix $W$, replacing the diagonal elements by their absolute values, and then applying the inverse transformation back to the original basis. It does not correspond to taking the absolute value of each matrix element of $W$. Note that if we replace $W$ by $|W|$, then ${\rm Tr}KW^2$ is unchanged, because $W^2=|W|^2$. For the second term, we expand the trace in the basis where $W$ and $|W|$ are both diagonal. In that basis, we have both $W$ and $|W|$ are diagonal with matrix elements $w_\alpha$ and $|w_\alpha|$, respectively. We find ${\rm Tr}WL_xWL_x=\sum_{\alpha\beta}w_\alpha w_\beta |(L_x)_{\alpha\beta}|^2$, while ${\rm Tr}|W|L_x|W|L_x=\sum_{\alpha\beta}|w_\alpha||w_\beta||(L_x)_{\alpha\beta}|^2$. Hence, because $U<0$, the energy $E$ cannot increase when we compute the variational energy for the state $|W|$. This means that if $W$ is a ground state, then $|W|$ is also a ground state. This result is called spin-reflection positivity.

So amongst all possible ground states in the $S_z=0$ sector, at least one satisfies $W=|W|$. It turns out that this condition implies that the ground state includes a spin singlet with $S=0$. To prove this, we need to show that at least one diagonal element of $|W|$ is nonzero when we express the wavefunction in the localized basis (corresponding to product states of $c_{x}^\dagger$ acting on the vacuum state). This is because a nonzero diagonal matrix element means that the wavefunction has a nonzero coefficient for the state that has all of the up spins at precisely the same lattice sites as the down spins. This state is a spin singlet because all doubly occupied sites are spin singlets. The proof is simple. The matrix $|W|$ is positive semidefinite. But all positive semidefinite matrices must have diagonal elements that are positive or zero and they must have at least one nonzero diagonal element or they are identically zero. But since ${\rm Tr}|W|^2=1$, we cannot have $|W|=0$, so at least one diagonal element is nonzero. This means amongst all of the ground states, at least one is a spin singlet.

Next, Lieb showed that if the lattice is connected, then the ground state is unique. We now summarize that proof. The basic strategy is to show that if $W$ is a ground-state, then we must have $W=\pm|W|$. Suppose there are two ground-states $W_1$ and $W_2$. Then the matrix $W(\alpha)=W_1+\alpha W_2$ is also a ground-state. If we pick $\alpha=-{\rm Tr}W_1/{\rm Tr}W_2$, then ${\rm Tr}W(\alpha)=0={\rm Tr}|W(\alpha)|$. This means that the ground-state corresponding to $W(\alpha)$ is neither positive semidefinite nor negative semidefinite. But, since $|W(\alpha)|$ is also a ground-state and is positive semidefinite with a vanishing trace, it must be identically zero, which is not possible. So, one cannot have two linearly independent ground-states. 

In order to prove that $W=\pm|W|$, we construct the kernel of the matrix $R=|W|-W$, which is positive semidefinite.
The kernel of $R$ is the set of vectors $\{V\}$ that satisfy $RV=0$. To establish this result, we need to show the kernel of $R$, which we denote by $Q$ is either the entire vector space, or just the null vector. We start from the Schr\"odinger equation in Eq.~(\ref{eq: schroedinger}), and take the expectation value with respect to a vector $V$ in $Q$:
\begin{equation}
V^\dagger(KR+RK+U\sum_{x\in\Lambda}L_xRL_x)V=EV^\dagger RV.
\label{eq: r1}
\end{equation}
But $RV=V^\dagger R=0$, so we have $\sum_{x\in\Lambda}V^\dagger L_xRL_xV=0$. This then implies that if $V\in Q$ then $L_xV\in Q$, because the positive semidefiniteness of $R$ implies that each $V^\dagger L_xRL_xV=0$ which requires $RL_xV=0$. Hence, $L_x$ maps the kernel $Q$ into itself. Now multiply the Schr\"odinger equation on the right by $V$, to find 
\begin{equation}
(KR+RK+U\sum_{x\in\Lambda}L_xRL_x)V=ERV.
\label{eq: r2}
\end{equation}
Since $RV=0$ and $RL_xV=0$, we find $RKV=0$. Hence if $V\in Q$, then $KV\in Q$. 

The hopping matrix $t_{xy}$ creates a bond between lattice sites $x$ and $y$ if $t_{xy}\ne 0$. The hopping matrix is said to be connected, if for any two lattice sites $x$ and $y$, there a set of bonds that connect $x$ with $y$. This can also be stated as there is a chain of lattice sites $x=x_0,\,x_1,\,x_2,\,\ldots,x_n,\,y=x_{n+1}$ such that $t_{xx_1}t_{x_1x_2}\cdots t_{x_ny}\ne 0$. The connectedness property translates from the hopping matrix to the many-body states in the localized basis introduced earlier. This means for any state $c_{x_1}^\dagger c_{x_2}^\dagger\cdots c_{x_M}^\dagger|0\rangle$ is connected to any other state $c_{y_1}^\dagger c_{y_2}^\dagger\cdots c_{y_M}^\dagger |0\rangle$ by a chain of $K_{\alpha\beta}$ and $L_x$ matrices. To find the nonzero chain of matrix elements, we first find a path in the single-spin many-body space that connects the state with a fermion at sites $x_1,\, x_2,\,\ldots ,\,x_M$ to $y_1,\, y_2,\,\ldots,\,y_M$. This path is made in the following fashion\cite{lieb_freericks}: Suppose that $x_M\ne y_M$. Construct a path with the hopping matrix elements from site $x_M$ to $y_M$ using nonzero hopping matrix elements. Such a path always exists because the hopping matrix is connected. Locate every fermion in the set $\{x_i\}$ that lie on the path. Starting with the last fermion in the chain (which could be the one at $x_M$ if no other fermions lie on the path), move it one step at a time from its current location until it reaches $y_M$. Next, take the next to last fermion on the path, and move it to the position occupied by the last fermion before it was moved. Continue in turn doing this for all fermions on the path until you have moved all of them, including the last one $x_M$. At this stage, we have moved the fermion from $x_M$ to $y_M$, and have left all of the other fermions in the set $\{x_i\}$ unchanged. Repeat this procedure for $x_{M-1}$ and $y_{M-1}$, for $x_{M-2}$ and $y_{M-2}$, all the way down to $x_1$ and $y_1$. Then we have found the path in the many-body space connecting these two states. Consider an elemental step where we start with a fermion at site $\alpha$ which we can move to site $\beta$ since $t_{\alpha\beta}\ne 0$; assume as well that all the other sites are labeled by $z_i$ for $i=1,\ldots,M-1$. The operator that connects these two many body states is $L_\beta(\prod_{x\in\{z_i\}}L_x)KL_\alpha(\prod_{x\in\{z_i\}}L_x)$, which is composed entirely by $K$ and $L_x$ matrices. Hence, the operator that connects the state with fermions at $\{x_i\}$ to the state with fermions at $\{y_i\}$ can be constructed entirely out of $K$ and $L_x$ matrices. Suppose that $Q$ has a nonzero vector $V$ in it. There is a basis state, where the fermions are at sites $\{x_i\}$, such that $(\prod_{x\in\{x_i\}}L_x)V\ne 0$. Then we can connect this state to any other many-body state using products of $K$ and $L_x$ matrices. Hence, every basis vector in the many-body space can be mapped to every other basis vector. Since we started with a vector that is in $Q$, and the $K$ and $L_x$ matrices keep vectors inside $Q$, we have shown that all vectors in the many-body space are in $Q$. The other option is that $Q$ has no vectors in it except for the zero vector. This then proves that $W=\pm|W|$, which then implies that the ground-state is unique. 

Without loss of generality, we assume $W=|W|$. With just a little more work we can show that the ground-state is actually positive definite not just positive semidefinite. The argument follows precisely what was given above, but we use the matrix $|W|$ instead of $R$. Suppose $|W|V=0$, so that $V$ is in the kernel of $|W|$. Then following the above steps, we immediately find that either the kernel of $|W|$ contains only the zero vector or it contains every vector. If it contains every vector, then $|W|=0$,which is not possible because ${\rm Tr}W^2=1$, so it must contain just the null vector. This implies that $|W|$ is actually positive definite. This extension of Lieb's theorem was made by Shen.

\section{Summary of Shen's pseudospin proof}

We prove a straightforward lemma from Shen that if one wavefunction is positive definite and the other is positive semidefinite, then their overlap is nonzero\cite{shen_review}. Let $W$ be the positive definite wavefunction and $W'$ the positive semidefinite wavefunction. Introduce the unitary matrix $U$ which diagonalizes $W$ via $W=U^\dagger D U$, where $D$ is a diagonal matrix whose diagonal elements all satisfy $|w_\alpha|>0$. Then, we can immediately compute the overlap
\begin{eqnarray}
{\rm Tr}W'^\dagger W&=&{\rm Tr}W'^\dagger U^\dagger D U={\rm Tr}UW'^\dagger U^\dagger D=\sum_{\alpha}(UW'^\dagger U^\dagger)_{\alpha\alpha}|w_\alpha|\nonumber\\
&=&\sum_{\alpha\beta\gamma}U_{\alpha\beta}W'_{\beta\gamma}U^*_{\alpha\gamma}|w_\alpha|.
\end{eqnarray}
But $\sum_{\beta\gamma}U_{\alpha\beta}W'_{\beta\gamma}U^*_{\alpha\gamma}\ge 0$ for every $\alpha$ because $W'$ is positive semidefinite. Not all can be zero or $W'=0$, which is not possible if ${\rm Tr}W'^\dagger W'=1$. Hence, the summation includes some nonzero terms and all nonzero terms are positive. Therefore, ${\rm Tr}W'^\dagger W>0$.

\begin{figure}[th]
\centerline{\psfig{file=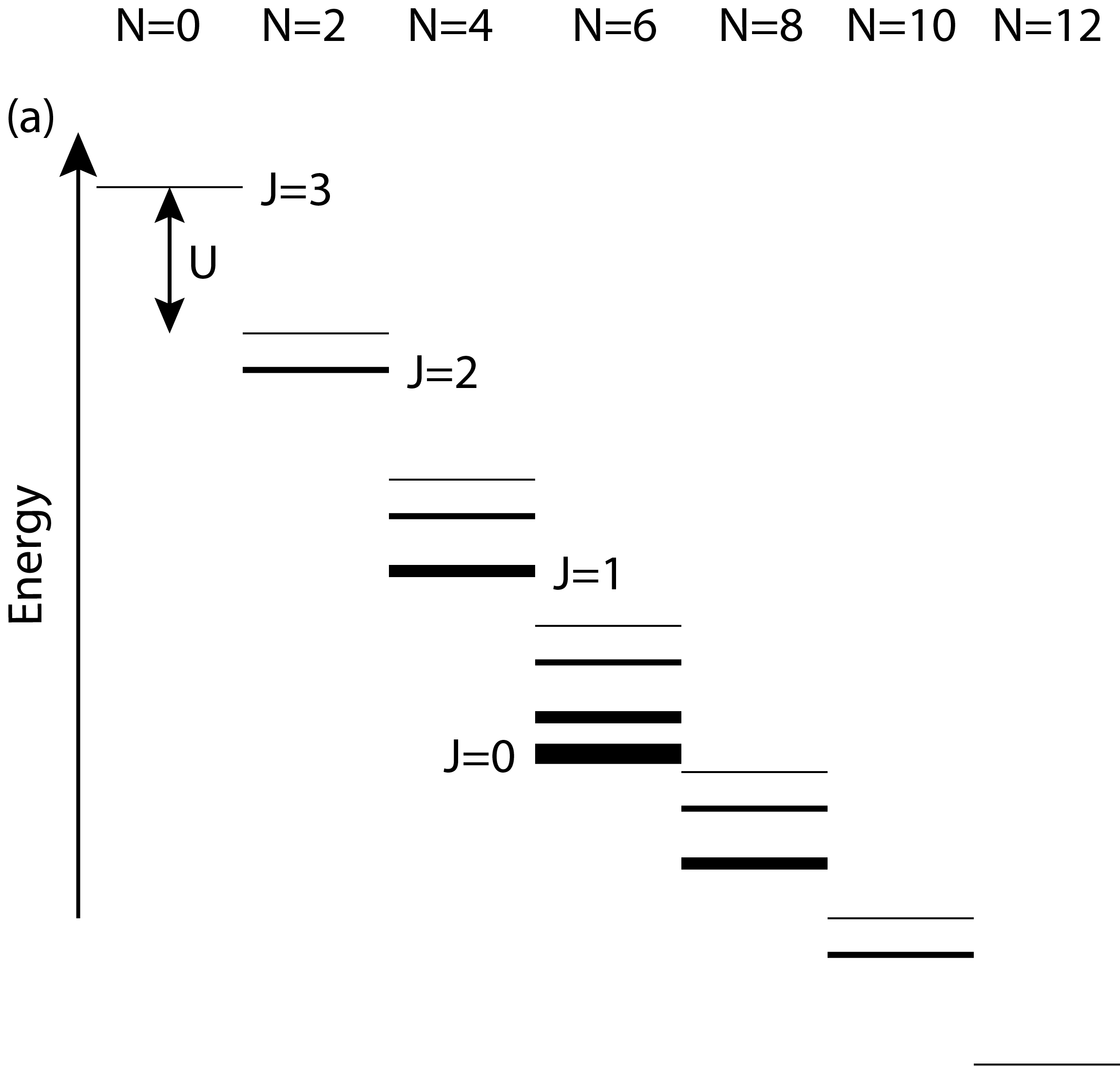,width=2.4in}~~~~
\psfig{file=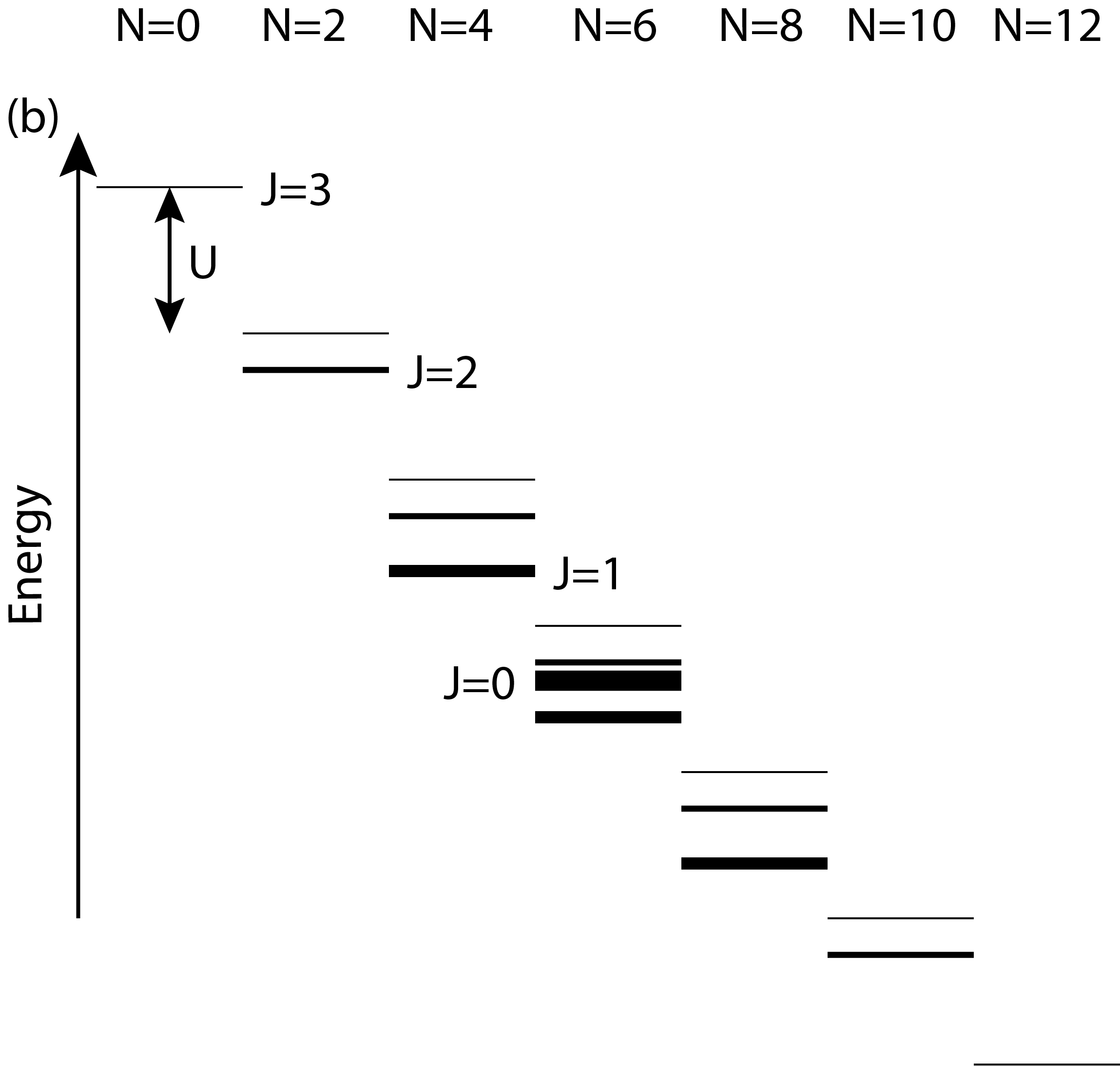,width=2.4in}}
\caption{Schematic of a pseudospin tower for the attractive Hubbard model. (a) Pseudospin tower for $|\Lambda|=6$ on a lattice with $N_A=N_B$ and (b) pseudospin tower for $|\Lambda|=6$ on a lattice with $N_A=2$ and $N_B=4$. The energy levels are plotted on the vertical axis and the electron number on the horizontal. We only show the lowest energy level for each pseudospin for a given number of electrons. The thickness of the line denotes the pseudospin value, as labeled in the figure. The energy decreases by exactly $-|U|$ as the electron number increases by 2, as shown in the figure. In panel (b), the minimal energy state has $J=1$ for 4, 6, and 8 electrons due to the partial pseudospin tower, which occurs on a Lieb lattice.\label{fig: pseudo}}
\end{figure}

A full pseudospin tower occurs for a fixed filling of electrons $N=2M$, when the minimal energy state for a given pseudospin is ordered according to the pseudospin values from the lowest allowed pseudospin $J=(N_B+N+A-2M)/2$ up to the maximal pseudospin $J=(N_B+N_A)/2$; that is $E_{min}(J+1)>E_{min}(J)$ for all $J\ge (N_B+N_A-2M)/2$. Here, the subscript $min$ denotes the minimal energy state with the given value of pseudospin.  A partial pseudospin tower has the minimal energy ordering down to a given pseudospin $J_0$ and for pseudospins with lower quantum numbers the minimal energy states are all larger than the energy of the minimal state with pseudospin $J_0$. These different cases are illustrated in Fig.~\ref{fig: pseudo}. For the attractive Hubbard model, we have a full pseudospin tower for the case where $N_A=N_B$, while we have $J_0=(N_B-N_A)/2$ for a Lieb lattice. Next, we develop how one can prove the existence of a pseudospin tower.

One can employ Shen's lemma to find the quantum number of the ground state of the attractive Hubbard model by finding any positive semidefinite state that has a definite value of the particular quantum number. Then since the overlap of the ground state with the positive semidefinite state is nonzero, it must share the same eigenvalue with the ground state. Shen did this by finding a specific pseudospin model that had positive semidefinite ground states, from which one can infer the overall quantum number of the ground state\cite{shen,shen_review}. We adopt the same methodology for determining the pseudospin of the ground state, but we next construct the positive semidefinite state directly rather than inferring it from the ground state of a pseudospin Hamiltonian.

\section{Direct construction of positive semidefinite pseudospin state}

The states we will consider consist of linear combinations of product states that involve pairs of up and down spin electrons created on specific lattice sites. When expressed in the matrix form using the localized basis, these states all are represented by diagonal matrices. If the diagonal elements are all nonnegative coefficients, then the state is a positive semidefinite state, which will have a nonzero overlap with the ground-state wavefunction. 

We begin with the case $M=0$ of no electrons.The zero-electron state has $J=N_A+N_B=|\Lambda|$. If we apply $J^+$ onto this state, then $J^+|0\rangle$ has coefficients that are negative and positive. The operator creates a linear combination of single paired electrons on each site---those on the A sublattice are multiplied by $-1$ and those on the B sublattice are multiplied by $+1$. Hence this state is not positive semidefinite. We can form an (unnormalized) positive semidefinite state by choosing every paired state on the A sublattice to have the coefficient $N_B$ and every paired state on the B sublattice to have the coefficient $N_A$
\begin{equation}
|\psi_+(N=2)\rangle=N_B\sum_{x\in\Lambda_A}c^\dagger_{x\uparrow}c^\dagger_{x\downarrow}|0\rangle+N_A\sum_{x\in\Lambda_B}c^\dagger_{x\uparrow}c^\dagger_{x\downarrow}|0\rangle.
\end{equation}
One can immediately show that $J^-|\psi_+(N=2)\rangle=0$, so that $J=(N_A+N_B-1)$ for this state, and hence the $N=2$ ground state has minimal pseudospin.


We can keep continuing in this fashion. The state with $N=2M$ electrons is composed of $M$ sites of paired electrons. We form the positive semidefinite state with the same coefficient for product states that have $m_A$ pairs on the $A$ sublattice and $m_B$ pairs on the $B$ sublattice ($M=m_A+m_B$), so they have the form
\begin{equation}
|\psi_+(N=2M)\rangle=\sum_{m_A=0}^M c_{m_A,M-m_A}\sum_{\substack{{\rm all~states~with~}\\m_A~{\rm pairs~in~}\Lambda_A{\rm ~and~}\\M-m_A{\rm ~pairs~in~}\Lambda_B}}|m_A;M-m_A\rangle,
\end{equation}
with all $c_{m_A,M-m_A}\ge 0$; in other words, each state of the same type has the same coefficient in the linear combination, but different types have different coefficients. The coefficients are chosen such that $J^-|\psi_+(N=2M)\rangle=0$, so that the state has $J=(N_A+N_B-M)$. Finding these coefficients is a simple counting exercise. For a given $m_A$, we have $N_A!/[m_A!(N_A-m_A)!]\times N_B!/[(M-m_A)!(N_B-M+m_A)!]$ different states. When $J^-$ acts on this state, each term on the $A$ sublattice will create $m_A$ new terms of the form $-|m_A-1;M-m_A\rangle$ and $M-m_A$ terms of the form $|m_A,M-m_A-1\rangle$. We need the negative terms with the same numbers of pairs on each sublattice to cancel against the positive terms. The total number of negative terms is $m_AN_A!/[m_A!(N_A-m_A)!]\times N_B!/[(M-m_A)!(N_B-M+m_A)!]$. This is equal to $(N_A-m_A+1)$ copies of the $N_A!/[(m_A-a)!(N_A-m+1)!]\times N_B!/[(M-m_A)!(N_B-M+m_A)!]$ terms in the $|m_A-1;M-m_A\rangle$ sector. Since each individual term in the set of $|m_A-1;M-m_A\rangle$ set of states is created from $N_A-m_a+1$ possible ``father'' states when a pair is removed from the $A$ sublattice and $N_B-M+m_A$ ``father'' states when a pair is removed from the $B$ sublattice, all terms appear the same number of times. This implies
\begin{equation}
- (N_A-m_A+1) c_{m_A,M-m_A}
+(N_B-M+m_A)c_{m_A-1,M-m_A+1}=0,
\end{equation}
for $1\le m_A\le M$ when $M< N_A$. Start with $c_{0,M}$. Then $c_{1,M-1}=(N_B-M+1)c_{0,M}/N_A$, $c_{2,M-2}=(N_B-M+1)(N_B-M+2)c_{0,M}/[N_A(N_A-1)]$, $\cdots$, 
\begin{equation}
c_{m,M-m}=\frac{(N_B-M+m)!}{(N_B-M)!}\frac{(N_A-m)!}{N_A!}c_{0,M}.
\label{eq: coeffs}
\end{equation}
As long as $M<N_A$, then this state $|\psi_+(N=2M)\rangle$ does not remain positive semidefinite when $J^+$ is applied to it, because all of the states with $M+1$ pairs on the $A$ sublattice and none on the $B$ sublattice have negative coefficients coming from the terms in $J^+$ which create pairs on the $A$ sublattice and acted on the states of the form $|M;0\rangle$. Hence, as long as $M<N_A$, the ground state has minimal pseudospin.

When $M=N_A$, we cannot add any new pairs to the $A$ sublattice from the $|N_A=M,0\rangle$ state, so it may be possible now that applying $J^+$ to this state can lead to a positive semidefinite state. In fact, by restricting the coefficients in Eq.~(\ref{eq: coeffs}) to run only over $1\le m_A\le N_A$ for $N_A\le M\le N_B$,
we find that the $J^+$ operator can be applied $N_B-N_A$ times and the coefficients remain all positive; they are multiplied by $N_B-M$ for $N_A\le M\le N_B$. So the states with $2N_A\le N\le 2N_B$ all have $J=N_B-N_A$ and the ground-state pseudospin is no longer minimal.

Finally, as $M$ runs from $N_B$ to $N_A+N_B$, the ground state is minimal pseudospin again. This can be seen because the positive-definite state with a definite $J$ quantum number no longer remains positive definite when $J^+$ is applied to it. This occurs because the state in the sector $|M-N_B;N_B\rangle$ has negative coefficients for all the $M-N_B+1;N_B\rangle$ states and since they all come from just the $|M-N_B;N_B\rangle$ sector, the states are not positive semidefinite. This implies that any positive semidefinite state with a definite $J$ value  cannot be raised and still remain positive semidefinite. Hence the positive semidefinite state must be annihilated when $J^+$ is applied, implying the pseudospin is minimal.

This establishes the proof about the pseudospin quantum number for the ground state of the attractive Hubbard model. Our proof employed a constructive method which created positive-semidefinite states with definite $J$ which have nonzero overlap with the ground state and hence share the same quantum number.

\section{Shen's proof of the spin tower}

All of the work we have done so far was for the attractive case with $U<0$. The repulsive case can be connected to the attractive one via a partial particle-hole transformation, where we perform a particle-hole transformation on the down spin electrons\cite{lieb}. This takes $c_{x\downarrow}^{\phantom{\dagger}}\rightarrow (-1)^{\Pi_x}c_{x\downarrow}^\dagger$ and $c^\dagger_{x\downarrow}\rightarrow (-1)^{\Pi_x}c^{\phantom{\dagger}}_{x\downarrow}$ but leaves the up spin electron operator unchanged. Then we find that the spin operators transform to the pseudospin operators and {\it vice versa}. The down spin filling is transformed from $M$ electrons to $|\Lambda|-M$ electrons. Finally, the interaction is changed via $U\rightarrow -U$.
This transformation changes none of the anticommutation relations.

Let's examine what happens to the $S_z=0$ states with $N=2M$ electrons ($M$ up spin and $M$ down spin). The electron filling changes to $M$ up spins and $|\Lambda|-M$ down spins, so the total filling is $N=|\Lambda|$, which is what we call half-filling. The $z$ component of the spin is now $M-|\Lambda|/2$, so as $M$ runs from 0 to $|\Lambda|$, $S_z$ runs from $-|\Lambda|/2$ to $|\Lambda|/2$.

\begin{figure}[th]
\centerline{\psfig{file=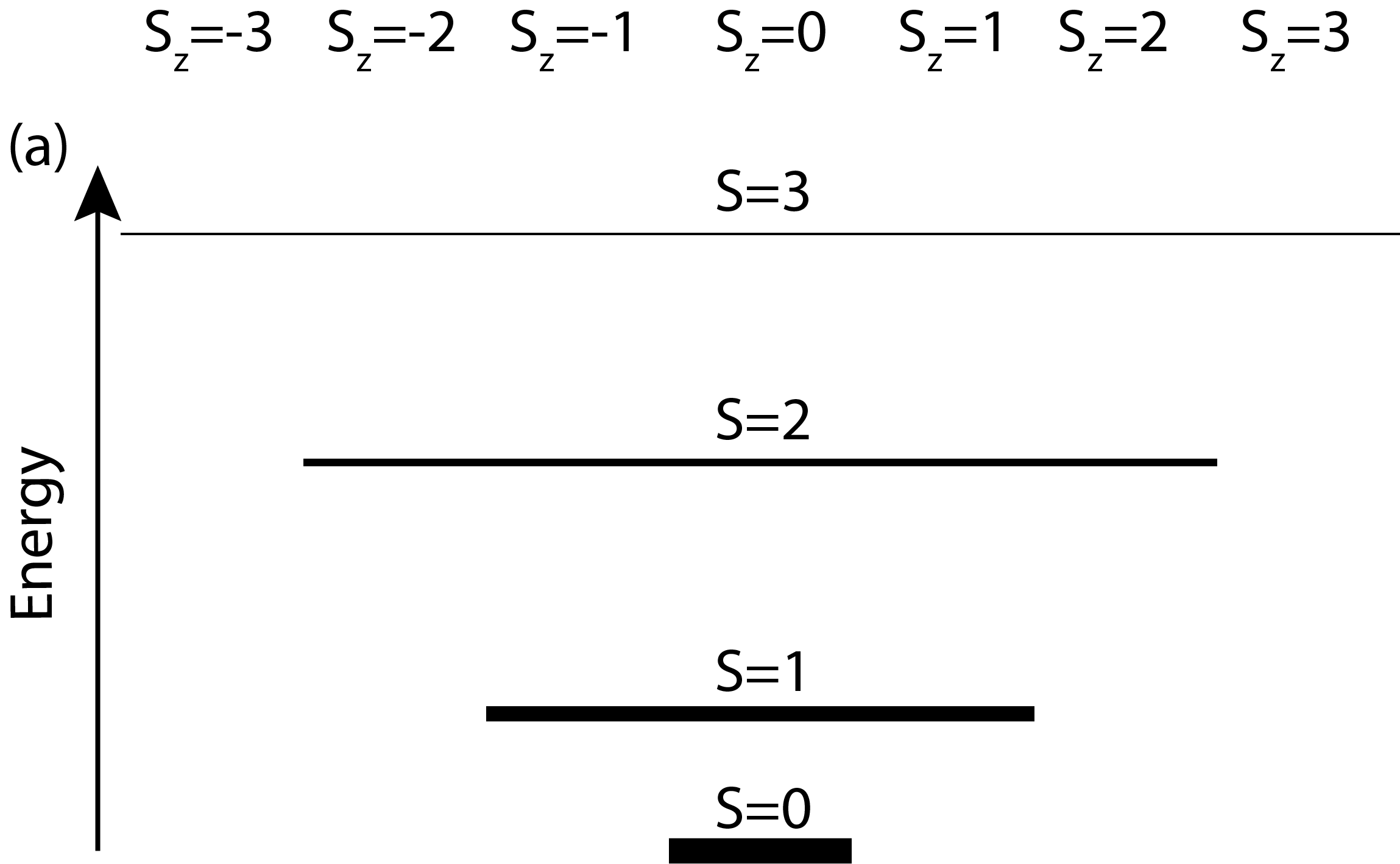,width=2.4in}~~~~
\psfig{file=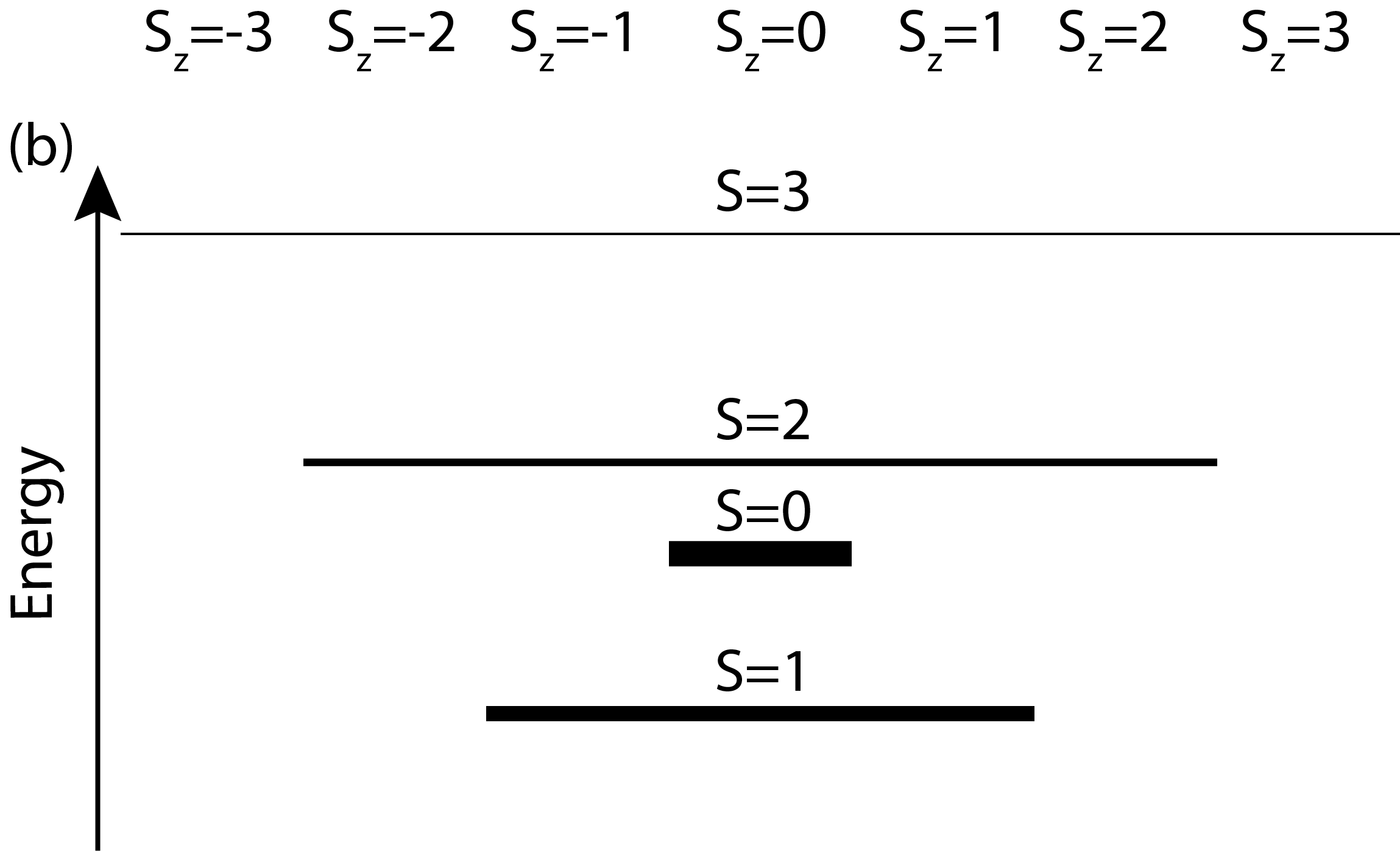,width=2.4in}}
\caption{Schematic of a spin tower at half-filling for the repulsive Hubbard model. (a) Spin tower for $|\Lambda|=6$ on a lattice with $N_A=N_B$ and (b) spin tower for $|\Lambda|=6$ on a lattice with $N_A=2$ and $N_B=4$. The energy levels are plotted on the vertical axis and the $z$-component of the electron spin on the horizontal. We only show the lowest energy level for each total spin quantm number. The thickness of the line denotes the spin value, as labeled in the figure.  In panel (b), the minimal energy state has $S=1$ for $S_z=0,\pm 1$ due to the partial spin tower, which occurs on a Lieb lattice.\label{fig: spin}}
\end{figure}

If we plot the energy levels for half-filling as a function of $S_z$, a full spin tower implies that $E(|S_z|)>E(|S'_z|)$ whenever $|S_z|>|S'_z|$. A partial spin tower is a spin tower for all $|S_z|>S_0$ and then $E(|S_z|)>E(S_0)$ for $|S_Z|<S_0$. This implies that the ground state has spin $S_0$ and a spin tower exists for spins larger than $S_0$. Shen proved that on a bipartite lattice with $N_A=N_B$, we have a full spin tower and the ground state is a spin singlet at half-filling for the repulsive case when $U<\infty$. This result generalizes the Lieb-Mattis tower, which was proved for attractive and repulsive interactions for one-dimensional systems. Shen also showed that $S_0=(N_B-N_A)/2$ for Lieb lattices. The situation is shown schematically in Fig.~\ref{fig: spin}.

The proof is rather straightforward\cite{shen,shen_review}. We first need to relate the energies of the attractive and repulsive cases. In the current Hamiltonian, we have the interaction energy is $U\sum_{x\in\Lambda}c^\dagger_{x\uparrow}c^{\phantom{\dagger}}_{x\uparrow}
c^\dagger_{x\downarrow}c^{\phantom{\dagger}}_{x\downarrow}$, which clearly gives a different result if 
we have zero, one or two electrons at a site. But if we instead write it in the particle-hole symmetric form as
$U\sum_{x\in\Lambda}(c^\dagger_{x\uparrow}c^{\phantom{\dagger}}_{x\uparrow}-\frac{1}{2})
(c^\dagger_{x\downarrow}c^{\phantom{\dagger}}_{x\downarrow}-\frac{1}{2})$, then the interaction is the same when there are zero or two electrons on a site, while the case with one electron on a site gives the negative value. Since the partial particle-hole transformation changes the down filling from one to zero and {\it vice versa} as well as changing the sign of $U$, it actually does not change the potential energy when written in this form. This means the attractive and repulsive cases have the same potential energy. The kinetic energy is also unchanged for the half-filled case, because the kinetic energy for the up spins are unchanged in the transformation. For the down spins, the filling goes from $M$ to $|\Lambda|-M$, but because of the bipartite nature of the hopping, the kinetic energy with $M$ particles and with $M$ holes (or $|\Lambda|-M$ particles) is identical. Hence the energies are the same for the modified interaction after the particle-hole transformation.  

Now the conversion from the original interaction and the particle-hole symmetric one requires us to add $-U\sum_{x\in\Lambda}(c^\dagger_{x\uparrow}c^{\phantom{\dagger}}_{x\uparrow}+
c^\dagger_{x\downarrow}c^{\phantom{\dagger}}_{x\downarrow})+U|\Lambda|/4$. All of our previous energy eigenstates are eigenstates of this additional operator as well, so the shift of the eigenvalues can be immediately computed. More importantly, the shift has no effect on the pseudospin eigenvalues or the spin eigenvalues. 

So, at this stage we know the following: (1) the $S_z=0$ attractive eigenstates map to $J_z=0$ repulsive eigenstates corresponding to half-filling; (2) for a given $J_z$ value for the attractive case, we know the $J$ value of the minimal energy eigenstate, so for the repulsive case at half-filling we know the $S$ eigenvalue of the minimal energy state for fixed $S_z$; and (3) the energy eigenvalues for the repulsive and attractive cases are the same. 

Now focus on the repulsive case at half-filling. We have that $S_0=(N_B-N_A)/2$. If $S_z>S_0$, then we know the minimal energy states for $|S_z|>|S'_z|$ must satisfy $E(|S_z|)>E(|S'_z|)$. This follows because in the attractive case, when we apply $J^+$ to the minimal energy state with a given $J_z<0$ value, the energy changes by $U-U=0$ for the modified potential. But the minimal energy state with $J_z+1$ must be lower in energy. Performing the partial particle-hole transformation then yields the desired result. If $|S_z|< S_0$, then $E(|S_z|)>E(S_0)$ follows because the minimal energy state is a pseudospin multiplet for the attractive case. This then establishes the (partial) spin tower. Since we are at half-filling, the term we added to the Hamiltonian is just a constant, so we can remove it and go back to the original Hamiltonian, and the result remains the same---the repulsive Hubbard model has the same spin tower at half-filling.

Note that this shows that the ground state is ferrimagnetic (with $S=|N_B-N_A|/2$) on a Lieb lattice. In particular, it agrees with the known spin of the ground state of an antiferromagnetic Heisenberg model, to which the repulsive Hubbard model maps when $U$ is positive and large in magnitude\cite{lieb_mattis,lieb}.

\section{Remaining conjecture about ground-state quantum numbers and towers for the Hubbard model}

What remains to be proved is that the attractive case has a spin tower for all even fillings. If true, then the particle-hole transformation would show that the pseudospin quantum number of the repulsive model with even numbers of electrons has minimal pseudospin. Heuristically, this result is obviously true when $U$ is negative and large in magnitude. For then it costs too much energy to unbind a pair of electrons and there is not enough kinetic energy gain, so the ground state has all electrons paired. Increasing the spin requires the breaking of a pair, which increases the energy by $|U|$, which occurs for each increased spin. Hence, one gets a spin tower. The same thing occurs for small $U$ approaching zero, if the bandstructure is nondegenerate. Then, by filling the states in from the lowest energy level upward, we find we have to move pairs of electrons off the same energy level to higher levels, which also will yield the spin tower.

In general, it is difficult to use any of the similar strategies that worked with the above proofs for these cases because $N_\uparrow\ne N_\downarrow$ implies that the matrix $W$ is not square. Furthermore, because the up spin and down spin basis functions are different, we can no longer show that $W$ is Hermitian. The situation of nonzero $S_z$ and half-filling, is however, unique in that this case does have square matrices. Furthermore, if one can establish a spin tower here, one can establish it at all other fillings too by using pseudospin raising and lowering operators. But one cannot show that $W$ is Hermitian anymore, which makes the general approach to solving the problem difficult (that is, spin-reflection positivity cannot be employed in this case).

We spent some time focusing on singular value decompositions to employ with variational arguments, but were not able to be successful with any of these arguments. We feel the best line of attack at this point is to focus on different variational ideas that go beyond using the matrix representation for the wavefunction. But we do not have any concrete ideas for what might actually work.

\section{Conclusions}

In this work, we gave a brief review of Lieb's two theorems proved in 1989 about the Hubbard model and some of Shen's extensions of these proofs. In particular, we showed an alternative proof to Shen's proof of the pseudospin quantum number of the ground state for the attractive model and how one can employ that to establish a spin tower for the repulsive Hubbard model at half-filling.  Our proof involved a constructive approach for positive semidefinite wavefunction with definite values of pseudospin. We ended with some conjectures about the remaining open problem which would establish a spin tower for the attractive case and would determine the minimal pseudospin of the repulsive model ground state. Knowing the quantum number of the ground state of the Hubbard model could then be employed to improve exact diagonalization and other numerical methods for computing properties of these models.

\section*{Acknowledgments}

J.Z.B. was supported by the National Science Foundation under grant number DMR-1659532.
J.Z.B., J.K.F., and J.R.C. were supported by the National Science Foundation under grant number PHY-1620555. J.K.F. was also supported by the McDevitt bequest at Georgetown University.

\end{document}